\documentclass[Physsubmission, Phys]{SciPost}

\hypersetup{
    colorlinks,
    linkcolor={red!50!black},
    citecolor={blue!50!black},
    urlcolor={blue!80!black}
}

\usepackage[bitstream-charter]{mathdesign}
\newcommand{\sfac}{\mathfrak{s}}
\def\rQCED{{\rm QCED}}
\newcommand{\KK}{${\cal KK}$}

\DeclareSymbolFont{usualmathcal}{OMS}{cmsy}{m}{n}
\DeclareSymbolFontAlphabet{\mathcal}{usualmathcal}
\DeclareMathOperator{\sgn}{sgn}
\begin{document}

\begin{center}{\Large \textbf{
IR-Improved Amplitude-Based Resummation in Quantum Field Theory: New Results and New Issues\\
}}\end{center}

\begin{center}
B.F.L. Ward\textsuperscript{1$\star$},
S. Jadach\textsuperscript{2},
W. Placzek\textsuperscript{3},
M. Skrzypek\textsuperscript{2},
Z. Was\textsuperscript{2} and
S.A. Yost\textsuperscript{4}
\end{center}

\begin{center}
{\bf 1} Baylor University, Waco, TX, USA
\\
{\bf 2} Institute of Nuclear Physics, Krakow, Poland
\\
{\bf 3}  Institute of Applied Computer Science, Jagiellonian University, Krakow, Poland
\\
{\bf 4} The Citadel, Charleston, SC, USA
\\
* bfl\_ward@baylor.edu
\end{center}

\begin{center}
\today
\end{center}
\centerline{BU-HEPP-21-05}

\definecolor{palegray}{gray}{0.95}
\begin{center}
\colorbox{palegray}{
  \begin{tabular}{rr}
  \begin{minipage}{0.1\textwidth}
    \includegraphics[width=35mm]{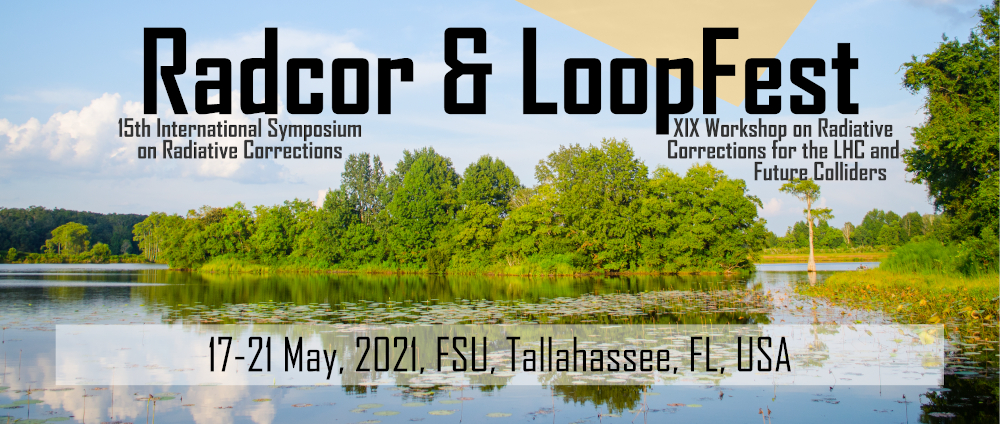}
  \end{minipage}
  &
  \begin{minipage}{0.85\textwidth}
    \begin{center}
    {\it 15th International Symposium on Radiative Corrections: \\Applications of Quantum Field Theory to Phenomenology,}\\
    {\it FSU, Tallahasse, FL, USA, 17-21 May 2021} \\
    \doi{10.21468/SciPostPhysProc.?}\\
    \end{center}
  \end{minipage}
\end{tabular}
}
\end{center}

\section*{Abstract}
{\bf
With the advancement of strategies for the precision physics programs for the HL-LHC, FCC-ee, FCC-hh, ILC, CLIC, CEPC, and CPPC, the need for proper control of the attendant theoretical precision tags is manifest. We discuss the role that amplitude-based resummation may play in this regard with examples from the LHC, the proposed new colliders and quantum gravity.

}

\vspace{10pt}
\noindent\rule{\textwidth}{1pt}
\tableofcontents\thispagestyle{fancy}
\noindent\rule{\textwidth}{1pt}
\vspace{10pt}

\section{Introduction}
\label{sec:intro}
\indent The subject of resummation in quantum field theory has applicability in the infrared (IR), collinear (CL), and ultraviolet (UV) limits. In what follows, we focus on the IR limit.
Independent of the limit in which one works, one can distinguish two general types of resummation: exact resummation and approximate resummation. Inside these two general classes, one has further sub-classes as we will illustrate. We start by defining the two general classes.\par
Toward this end, we recall that, even in elementary examples of summation, such as  $\sum_{n=0}^{\infty}x^n=1/(1-x)$, infinite order summation can greatly improve the behavior of the series relative to what one concludes from an examination of the series term-by-term. With this motivational background, we note that, in resummation, where we resum existing infinite order sums in quantum field theory, we can work either at the level of the Feynman amplitudes or at the level of observable cross sections. In either case, we can distinguish two types of resummation of the attendant Feynman series as follows:
 \begin{equation}
\sum_{n=0}^{\infty}C_n(Q,m)\tilde{\alpha}^n =\begin{cases} F_{\text{ RES}}(Q,m,\tilde{\alpha})\sum_{n=0}^{\infty}B_n(Q,m)\tilde{\alpha}^n, {\text{EXACT}}\\
                                                                                                      G_{\text{ RES}}(Q,m,\tilde{\alpha})\sum_{n=0}^{N}\tilde{B}_n(Q,m)\tilde{\alpha}^n, {\text {APPROX}},  
                                                                                                   \end{cases}
\label{eqres1}
\end{equation}
where $\tilde{\alpha}$ is the coupling constant expansion parameter (it is $\alpha_s$ in QCD, for example) and $N$ characterizes the order of the exactness of the approximate case labeled as 'APPROX' here while the first result on the right-hand side of the last equation defines what we call 'EXACT' resummation. $Q$ and $m$ are  generic representatives of momentum-dependent variables and masses that may enter the respective amplitudes or cross sections. As we have described in Ref.~\cite{radcor2019}, there is a long history of the comparison and competition between these two classes of resummation in quantum field theory. Generally,
there is a limit to the theoretical precision tag associated with the APPROX result determined by $N$ while, in principle, the EXACT result has no such limit.
In what follows, we use the EXACT representation of resummation in new paradigms to probe new issues in precision LHC/FCC physics and in quantum gravity.\par
As it has been noted in Refs.~\cite{case-sm50,slacsi2018}, we have recently passed 50 years of the Standard Theory\cite{SM3,SM4,SM1,BW1a,BW1b,BW1c}\footnote{We follow Prof. D.J. Gross~\cite{djg-smat50} and henceforth refer to the Standard Model as the Standard Theory of elementary particles.} of elementary particles. Given the current status of the observations at the LHC~\cite{susylmt1,susylmt2}, one has to keep a proper historical perspective with an eye toward the future. In Ref.~\cite{fcc-fabiola-2020}, one can see a path forward led by experiments at future energy frontier colliders (The interplay between experiment and theory is crucial to the progress of physics.). This allows us to look to the future with optimism.\par
The discussion is organized as follows. In the next Section, we review the representation of EXACT resummation which we will use. The discussion of its application to obtain new results with new issues in precision LHC physics is given in Section 3. The analogous discussion for the FCC is given in Section 4. Section 5 discusses
quantum gravity in this context. Section 6 contains our concluding remarks. \par

\section{Review of Exact Amplitude-Based Resummation Theory}
\label{sec:revw}
We include here a brief review of exact amplitude-based resummation theory, as it is still not generally familiar. The theory is encoded by the following master formula:
{\small
\begin{eqnarray}
&d\bar\sigma_{\rm res} = e^{\rm SUM_{IR}(QCED)}
   \sum_{{n,m}=0}^\infty\frac{1}{n!m!}\int\prod_{j_1=1}^n\frac{d^3k_{j_1}}{k_{j_1}} \cr
&\prod_{j_2=1}^m\frac{d^3{k'}_{j_2}}{{k'}_{j_2}}
\int\frac{d^4y}{(2\pi)^4}e^{iy\cdot(p_1+q_1-p_2-q_2-\sum k_{j_1}-\sum {k'}_{j_2})+
D_\rQCED} \cr
&{\tilde{\bar\beta}_{n,m}(k_1,\ldots,k_n;k'_1,\ldots,k'_m)}\frac{d^3p_2}{p_2^{\,0}}\frac{d^3q_2}{q_2^{\,0}},
\label{subp15b}
\end{eqnarray}}
where the {\em new}\footnote{The {\em non-Abelian} nature of QCD requires a new treatment of the corresponding part of the IR limit.}(YFS-style) residuals   
{$\tilde{\bar\beta}_{n,m}(k_1,\ldots,k_n;k'_1,\ldots,k'_m)$} have {$n$} hard gluons and {$m$} hard photons. The new residuals and the  infrared functions ${\rm SUM_{IR}(QCED)}$ and ${ D_\rQCED}$ are defined in Ref.~\cite{mcnlo-hwiri,mcnlo-hwiri1}.  Parton shower/ME matching leads to the replacements {$\tilde{\bar\beta}_{n,m}\rightarrow \hat{\tilde{\bar\beta}}_{n,m}$}, which allow us to connect with  MC@NLO~\cite{mcnlo,mcnlo1}, via the basic formula{\small
\begin{equation}
{d\sigma} =\sum_{i,j}\int dx_1dx_2{F_i(x_1)F_j(x_2)} d\hat\sigma_{\rm res}(x_1x_2s),
\label{bscfrla}
\end{equation}}
as explained in Ref.~\cite{mcnlo-hwiri,mcnlo-hwiri1}.\par
We have used Eq.(\ref{subp15b}) to obtain results in  precision LHC and FCC physics and we have extended it to general relativity as an approach to quantum gravity.  In each respective application, our new results are accompanied with new issues. We discuss such new results and new issues in precision LHC physics in the next Section.\par 

\section{Precision LHC Physics: New Results and New Issues}
\label{sec:lhc}
The large data sample of Z's and W's at the LHC affords the opportunity for precision EW studies as evidence by the ATLAS state-of-the-art measurement 
of $M_W$ in Ref.~\cite{atlasmw-17}. The effort to make an analogous state-of-the-art measurement of the weak mixing angle via $\sin^2\theta_W^{eff}$
is in progress in Ref.~\cite{froid:2019}. In this context, four of us (SJ, BFLW, SAY, ZW) have introduced the MC {\KK}MC-hh~\cite{kkmchh,kkmchh1,kkmchh2} which realizes exact ${\cal O}(\alpha^2 L)$
CEEX~\cite{ceex1:1999,ceex2:1999} EW corrections for hadron-hadron scattering processes with built-in Herwig6.5 and Herwiri1.031 showers as well as with an LHE\cite{lhe-formt} format for interfacing to other parton shower MC's. In the studies aimed at extracting observables such as $\sin^2\theta_W^{eff}$ from the precision analysis of the Z decays to lepton pairs at the LHC, the IR-improvement in {\KK}MC-hh opens the way for quantifying the effects of the ISR on such observables.\par
\begin{figure}[h]
\centering
\includegraphics[width=0.4\textwidth]{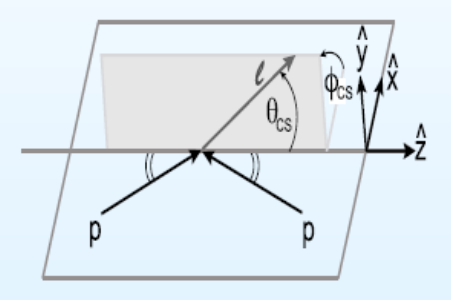}
\caption{Angular variables in the Collins-Soper frame.}
\label{fig1}
\end{figure}
\begin{figure}[h]
\centering
\includegraphics[width=0.7\textwidth]{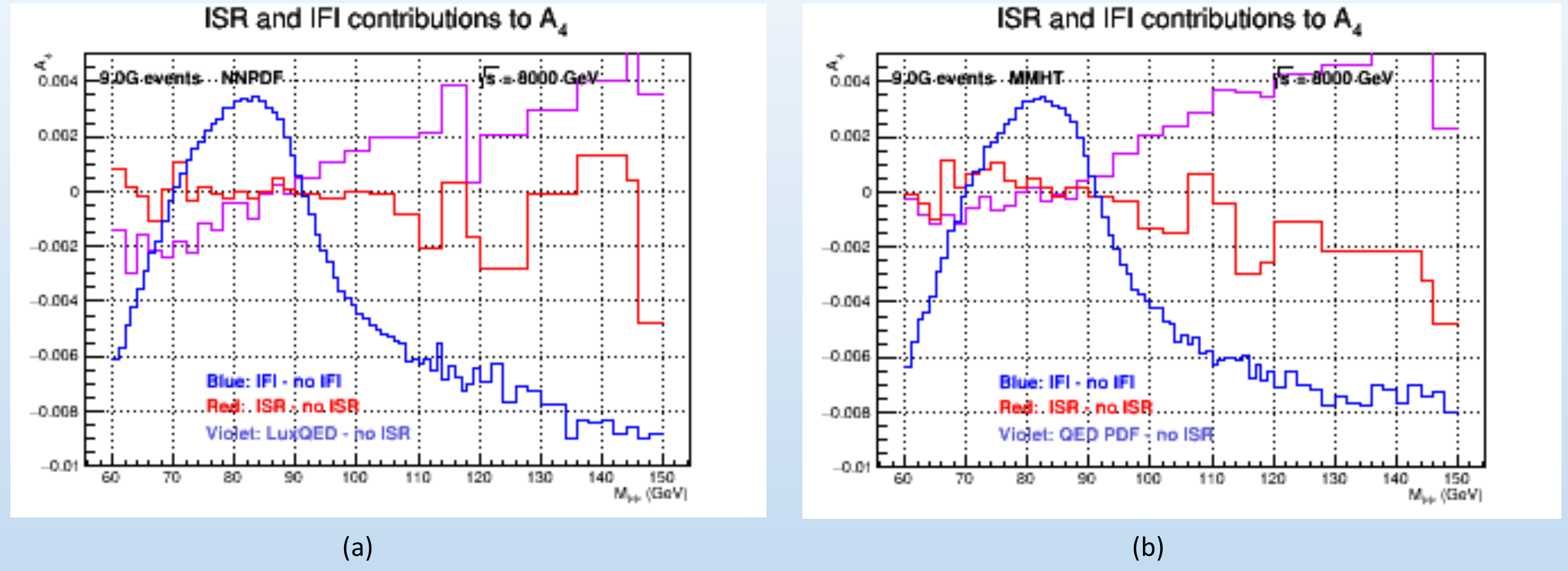}
\caption{PDF comparison for $A_4(\frac{8}{3}A_{FB})$} as a function of $M_{\ell\ell}$ with no lepton cuts: (a) shows results calculated with NNPDF3.1 NLO, (b) shows results calculated with MMHT2014 NLO. LUXQED denotes the QED-PDF realized by NNPDF3.1 and MMHT2014 following the approach of Ref.~\cite{luxqed}.
\label{fig2}
\end{figure}

We focus here on angular variables in $pp\rightarrow Z/\gamma^*\rightarrow \ell\bar{\ell}, \ell=e^-,\;\mu^-,$ augmented with the dilepton mass, $M_{\ell\ell}$, and rapidity, $Y_{\ell\ell},$ and consider distributions of the 
angle $\theta_{CS}$ of $\ell$ defined in the Collins-Soper~\cite{cs-frame} frame -- the CM frame of $\ell\bar{\ell}$, referenced to a $z$-axis oriented as shown in Fig.~\ref{fig1}.
 For $P=p_\ell+p_{\bar{\ell}},\; p^\pm=p^0\pm p^z$ in the Lab, we have
\begin{equation}
\cos(\theta_{CS})=sgn(P^z)\frac{p_\ell^+p_{\bar{\ell}}^- - p_\ell^-p_{\bar{\ell}}^+}{\sqrt{P^2P^+P^-}}.
\label{eqCS1}
\end{equation}
Using the angular distribution for $\theta_{CS}$, with an eye toward $\sin^2\theta_W^{eff}$, we may study, as we have done in Ref.~\cite{kkmchh2}, the observables $A_4$ and $A_{F B}$, 
which we define as 
$$A_4=\frac{4}{\sigma}\int\cos\theta_{CS}d\sigma=4<\cos\theta_{CS}>$$,
$$A_{FB}=\frac{1}{\sigma}\int \sgn(\cos\theta_{CS})d\sigma=<\sgn(\cos\theta_{CS})>$$
where we follow the notation of Ref.~\cite{mirkes-1} for the angular coefficients $A_i,\; i=0,\ldots, 7$ in the respective differential cross section $d\sigma(\theta_{CS},\phi_{CS})$ for the 
lepton in Fig.~\ref{fig1}. To illustrate our results, we start by showing in Fig.~\ref{fig2}, as a cross-check, the dependence of the size of the ISR and IFI effects on $A_4$, plotted as a function of $M_{\ell\ell}$, on the choice of PDF set when we feature results based on the NNPDF 3.1 NLO PDF set~\cite{nnpdf3.1} and on the MMHT2014 NLO set~\cite{mmht2014}. We see that the {\KK}MC-hh results with the two sets are consistent within the errors on the respective sets.
 In both cases, we see that the effects of the ISR are significant at the level of precision relevant for the expected uncertainty for the current LHC data analysis~\cite{froid:2019}. We also see from the difference between the ISR from {\KK}MChh and that from LUXQED that the transverse degrees of freedom in the photon radiation in {\KK}MChh are significant.\par
\begin{figure}[h]
\centering
\includegraphics[width=0.7\textwidth]{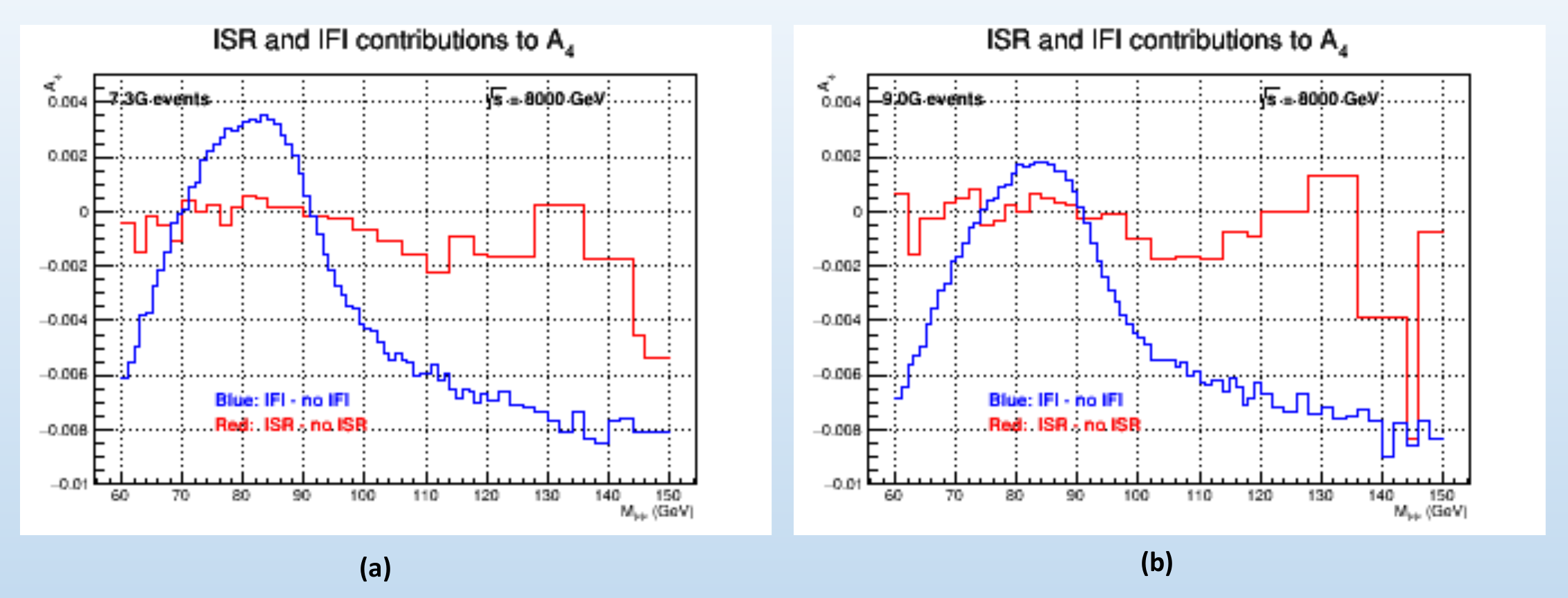}
\caption{Mass comparison for $A_4$} as a function of $M_{\ell\ell}$ with no lepton cuts: (a) shows results calculated with $m_u=2.2 MeV,\; m_d=4.7 MeV$, (b) shows results with these masses increased by a factor of 10.
\label{fig3}
\end{figure}

A new issue that our approach raises is the role of quark masses in precision LHC EW physics.  In this context, we show in Fig.~\ref{fig3}, using results from {\KK}MChh, 
the effect of changing the quark masses by a factor 10 on the size of the ISR and the IFI in $A_4 (\frac{8}{3}A_{FB})$, where the word 'heavy' in the figure refers to this factor of 10. The results, which again feature $A_4$ plotted as a function of $M_{\ell\ell}$
and are generated with the NNPDF3.1 NLO PDF set, show that the ISR is not very much affected by the change but that there is some effect on the IFI.\par
The role of the quark masses as observable parameters obtains in Figs.~\ref{fig3} -- they are not just a collinear regulators. This brings us the the issue of the input data at $Q_0\sim 1\;\text{GeV}$ for the non-QED PDFs used in {\KK}MChh. Is it a double counting if the collinearly singular quark mass terms are not removed? In local, relativistic quantum field theory, processes at different space-time regimes cannot double count one another. Thus, the quark mass singular terms in the input data at $Q_0\sim 1\;\text{GeV}$ do not double the quark mass singular terms at $Q\sim M_Z$; rather, the former produce an error in the probability to find the quark at the scale $Q\sim M_Z$, an error well below the error on the PDF itself.
Since it is possible~\cite{sjetaltoappear} to remove the quark mass singular terms from the input data at $Q_0\sim 1\;\text{GeV}$, it will soon be possible to remove this small error.\par

\section{Precision FCC Physics: New Results and New Issues}
\label{sec:fcc}
The FCC project received a strong endorsement from the CERN Council~\cite{fcc-fabiola-2020}. One possible scenario invovles the first stage with the FCC-ee for precision physics studies with more than 5 Tera Z's.
For such studies, the $e^+e^- $ luminosity theory error needs to be controlled at the precision of 0.01\% and five of us (SJ,WP,MS,BFLW,SAY) have shown~\cite{Jadach:2018} that, with sufficient resources, this theoretical precision tag
can be realized by upgrading the LEP era state-of-the-art MC BHLUMI4.04~\cite{bhlumi4:1996} from its current precision tag, which was recently shown to be $0.37\%$ in Ref.~\cite{pjnt-stj-2019}, to the desired 0.01\%. The steps required for this upgrade are discussed in detail Ref.~\cite{Jadach:2018}. For completeness, we show the resultant error budget for the upgraded BHLUMI in Table 1.
\begin{table}[ht!]
\centering
\scalebox{.85}{
\begin{tabular}{|l|l|l|l|}
\hline
Type of correction~/~Error
        & Update 2019
                &  FCC-ee forecast
\\ \hline 
(a) Photonic $[{\cal O}(L_e\alpha^2 )]\; {\cal O}(L_e^2\alpha^3)$
        & 0.027\%
                &  $ 0.1 \times 10^{-4} $
\\ 
(b) Photonic $[{\cal O}(L_e^3\alpha^3)]\; {\cal O}(L_e^4\alpha^4)$
        & 0.015\%
                & $ 0.6 \times 10^{-5} $
\\
(c) Vacuum polariz.
        & 0.013\%~\cite{JegerlehnerCERN:2016,fjeger-fccwksp2019}
                & $ 0.6 \times 10^{-4} $
\\
(d) Light pairs
        & 0.010\%~\cite{ON1,ON2}
                & $ 0.5 \times 10^{-4} $
\\
(e) $Z$ and $s$-channel $\gamma$ exchange
        & 0.090\%~\cite{BW13}
                & $ 0.1 \times 10^{-4} $
\\ 
(f) Up-down interference
    &0.009\%~\cite{Jadach:1990zf}
        & $ 0.1 \times 10^{-4} $
\\
(f) Technical Precision & (0.027)\% 
                & $ 0.1 \times 10^{-4} $
\\ \hline 
Total
        & 0.097\%
                & $ 1.0 \times 10^{-4} $
\\ \hline 
\end{tabular}}
\caption{\sf
Anticipated total (physical+technical) theoretical uncertainty 
for a FCC-ee luminosity calorimetric detector with
the angular range being $64$--$86\,$mrad (narrow), near the $Z$ peak.
Description of photonic corrections in square brackets is related to 
the 2nd column.
The total error is summed in quadrature.
}
\label{tab1}
\end{table}
\par
The results in the Table 1 and the discussion in Ref.~\cite{Jadach:2018} raise new issues and illustrate the synergies between the effort to realize the FCC-ee 0.01\% theoretical precision tag and 
other precision theory paradigms. For example, the need to realize the technical precision of $0.1 \times 10^{-4}$ means we will need two independent MC realizations that can realize the required precision. We can do this using the CEEX and EEX realizations of our YFS resummed MC methods or we can use an upgraded version of BabaYaga~\cite{babayaga-2019} or the further development of the results in Refs.~\cite{frixione-2019,bertone-2019,frixione-2021,signer:2021} to achieve the required cross check. The upgrade of BHLUMI for each of the items in Table ~\ref{tab1} is synergistic. Preparation for the item (a) ${\cal O}(L_e\alpha^2 )$ upgrade allowed, via crossing, the upgrade of the CEEX 2f production in {\KK}MC\cite{kkcpc:1999,Jadach:2013aha} which, combined with Herwig6.5~\cite{HERWIG}, has now been extended to Z production in hadron-hadron collisions in the MC {\KK}MChh~\cite{kkmchh}. Indeed, the need to extend CEEX to BHLUMI leads naturally to its extension to the other LEP era MC's BHWIDE~\cite{bhwide:1997}, YFSWW3~\cite{yfsww:1998,yfsww3:2001} together with KORALW\&YFSWW3~\cite{kandy-2001}, and YFSZZ~\cite{yfszz:1997},all but the last of which will be needed for the various precision measurements near the Z pole and in the WW production and reconstruction as discussed in Ref.~\cite{MSkrzypek-2020}. Indeed, the first step toward the extension of CEEX to WW production has been given in Ref.~\cite{jadach-wwceex:2019}, where we note that the contact with the usual Kleiss-Stirling~\cite{kleiss-stirling:1985} spinor-product-based photon helicity infrared factors in CEEX obtains via
\begin{equation}
ej^\mu_X(k_I) =eQ_X\theta_X\frac{2p^\mu_X}{2p_Xk_i} \rightarrow \sfac_{\sigma_i}(k_i)=eQ_X\theta_X\frac{b_{\sigma_i}(k_i,p_X)}{2p_Xk_i},
\end{equation}
with
\begin{equation}
b_{\sigma}(k,p_X)=\sqrt{2}\frac{\bar{u}_\sigma(k)\not p \mathfrak{u}_\sigma(\zeta)}{\bar{u}_{-\sigma}(k)\mathfrak{u}_\sigma(\zeta)},
\end{equation}
where the $\mathfrak{u}_\sigma(\zeta)$ are defined in Ref.~\cite{ceex2:1999}. The way forward is an open one.
\par
 
\section{Quantum Gravity: New Results and New Issues}
\label{sec:rqg}
With an eye toward the question of whether or not quantum gravity is calculable in relativistic quantum field theory,which is still open in the literature\footnote{See Refs.~\cite{rqg-ichep18,ijmpa2018} for further discussion on this point.}, we turn to the role of IR-improvement in quantum gravity. In this context, one of us (BFLW) argues that the attendant calculability holds
if he extends the YFS~\cite{yfs:1961,yfs1:1988,ceex2:1999} version\footnote{YFS-type soft resummation and its extension to quantum gravity was also worked-out by Weinberg in Ref.~\cite{sw-sftgrav}.} of the exact resummation example to resum the Feynman series for the Einstein-Hilbert Lagrangian for quantum gravity.  Indeed, very much in analogy with what we see in the elementary example discussed just before  Eq.(\ref{eqres1}), the resultant resummed theory, resummed quantum gravity (RQG), is very much better behaved in the UV compared to what one would estimate from that Feynman series.\par
One of us (BFLW) discussed many of the interesting consequences of RQG in Refs.~\cite{bw1rqg,bw2rqg,bw2arqg,bw2irqg,drkuniv}. Among these results, here he calls attention to
the prediction for the cosmological constant $\Lambda$ from RQG. Specifically, in Ref.~\cite{drkuniv}, he has shown that the RQG theory, taken together with the Planck scale inflationary~\cite{guth,linde} cosmology formulation in Refs.~\cite{reuter1,reuter2}\footnote{The authors in Ref.~\cite{sola1} also proposed the attendant 
choice of the scale $k\sim 1/t$ used in Refs.~\cite{reuter1,reuter2}.} from the 
asymptotic safety approach to quantum gravity in 
Refs.~\cite{reutera,laut,reuterb,reuter3,litim,litim1,litim2,perc,perc1,perc2,perc3,perc4}, allows him to predict, employing the arguments in Refs.~\cite{branch-zap}, the cosmological constant $\Lambda$ via the result {\small
\begin{equation}
\begin{split}
\rho_\Lambda(t_0)&\cong \frac{-M_{Pl}^4(1+c_{2,eff}k_{tr}^2/(360\pi M_{Pl}^2))^2}{64}\sum_j\frac{(-1)^Fn_j}{\rho_j^2}
          \times \frac{t_{tr}^2}{t_{eq}^2} \times (\frac{t_{eq}^{2/3}}{t_0^{2/3}})^3\cr
   & \cong \frac{-M_{Pl}^2(1.0362)^2(-9.194\times 10^{-3})}{64}\frac{(25)^2}{t_0^2}
   \cong (2.4\times 10^{-3}eV)^4.\cr
\end{split}
\label{eq-rho-expt}
\end{equation}}
$t_0\cong 13.7\times 10^9$ yrs  is the age of the universe, $t_{tr}\sim 25 t_{Pl}$ is the transition time between the Planck regime and the classical Friedmann-Robertson-Walker(FRW) regime in 
the Planck scale cosmology description of inflation in Ref.~\cite{reuter2}, $c_{2,eff}\cong  2.56\times 10^4$ is defined in Refs.~\cite{bw1rqg,bw2rqg,bw2arqg}, $t_{eq}$ is the time of matter-radiation equality, and $M_{Pl}$ is the Planck mass.\par
One of us (BFLW) has discussed~\cite{drkuniv,eh-consist} the reliability, consistency and implications of the prediction's closeness to the observed value~\cite{cosm1a,cosm1b,pdg2008}, $\rho_\Lambda(t_0)|_{\text{expt}}\cong ((2.37\pm 0.05)\times 10^{-3}eV)^4$.  He argues that its uncertainty is at the level of a factor of ${\cal O}(10)$. There follow constraints on susy GUT's as well~\cite{drkuniv}. This means that RQG is now rife for further confrontations with observation, such as that suggested in Ref.~\cite{calcagni:2019} in which the RQG prediction for the behavior of Newton's law at the Planck scale could be probed by appropriate observables.\par
\section{Summary}
\label{sec:sum}
The exact amplitude-level resummation of the IR regime of quantum field theory, coupled with exact results to a given order in an exact
re-arrangement of the original Feynman series, has wide applicability for precision phenomenology. We see that the range of this applicability is a broad one, spanning, as it does, from the current precision LHC physics to the futuristic FCC precision physics program and reliable estimates for the quantum theory of gravity by one of us (BFLW). The future of particle physics is intimately interwoven with continued progress on exact amplitude-level IR-improved (resummed) quantum field theoretic predictions, when taken together with control of the concurrent collinear and UV limits. More precise data require more precise theory.\par 

\section*{Acknowledgments}

This work was supported in part
by the Programme of the French–Polish Cooperation between IN2P3 and COPIN 
within the Collaborations Nos. 10-138 and 11-142 and by a grant from the Citadel Foundation. The authors also thank Prof. G. Giudice for the support and kind hospitality of the CERN TH Department.


\bibliography{Tauola_interface_design}

\nolinenumbers

\end{document}